\documentclass{PoS}

\usepackage{amsmath}
\usepackage{amssymb}

\usepackage{cite}

\usepackage{epsfig}

\newcommand{\be}{\begin{equation}}
\newcommand{\ee}{\end{equation}}
\newcommand{\bea}{\begin{eqnarray}}
\newcommand{\eea}{\end{eqnarray}}
\newcommand{\bean}{\begin{eqnarray*}}
\newcommand{\eean}{\end{eqnarray*}}

\newcommand{\bra}{\langle}
\newcommand{\ket}{\rangle}

\newcommand{\re}{{\rm Re}}
\newcommand{\im}{{\rm Im}}
\newcommand{\eps}{\epsilon}
\newcommand{\bit}{\begin{itemize}}
\newcommand{\eit}{\end{itemize}}
\newcommand{\bmp}[1]{\begin{minipage}{#1cm}}
\newcommand{\emp}{\end{minipage}}

\title{On complex Langevin dynamics and zeroes of the measure I: Formal proof and simple models }

\ShortTitle{On complex Langevin dynamics and zeroes of the measure I}

\author{\speaker{Gert Aarts}\\
        Department of Physics, College of Science, Swansea University, Swansea, United Kingdom\\
        E-mail: \email{g.aarts@swan.ac.uk}}

\author{Erhard Seiler\\
        Max-Planck-Institut f\"ur Physik (Werner-Heisenberg-Institut), M\"unchen, Germany\\
        E-mail: \email{ehs@mppmu.mpg.de}}

\author{D\'enes Sexty\\
        Department of Physics, Bergische Universit\"at Wuppertal, Wuppertal, Germany \\
        Inst.\ for Theoretical Physics, E\"otv\"os University,  Budapest, Hungary \\
        E-mail: \email{sexty@uni-wuppertal.de}}

\author{Ion-Olimpiu Stamatescu\\
        Institut f\"ur Theoretische Physik, Universit\"at Heidelberg, Heidelberg, Germany\\
        E-mail: \email{I.O.Stamatescu@thphys.uni-heidelberg.de}}

\abstract{
In the complex Langevin approach to lattice simulations at nonzero density, zeroes of the fermion determinant lead to a meromorphic drift and hence a need to revisit the theoretical derivation. We discuss how poles in the drift affect the formal justification of the approach and then explore the various potential issues in simple models, in a manner that is applicable to heavy dense and full QCD. 
}

\FullConference{34th annual International Symposium on Lattice Field Theory\\
                 24-30 July 2016\\
                 University of Southampton, UK}

\begin{document}

 \section{Introduction}
 
 Complex Langevin dynamics has solved the sign problem in a number of theories with a complex weight due to a nonzero chemical potential \cite{Aarts:2013lcm,Sexty:2014dxa}, including heavy dense QCD  \cite{Aarts:2008rr,Seiler:2012wz,Aarts:2016qrv}, and progress for QCD with lighter quarks is underway \cite{Sexty:2013ica,Aarts:2014bwa,Sinclair:2015kva}.
 However, the presence of the fermion determinant causes a theoretical problem, since the Langevin drift is then no longer holomorphic: zeroes of the determinant cause poles in the drift. In practice, this may lead to incorrect convergence, but not necessarily so
 \cite{Mollgaard:2013qra,Splittorff:2014zca,Nagata:2016alq}.
 Since the formal justification  \cite{Aarts:2009uq,Aarts:2011ax}  relies (among others) on holomorphicity a reconsideration of the derivation is required. This is sketched below. The interplay between the poles and the real and positive distribution sampled during the Langevin process turns out to be essential and this is studied in a sequence of models, with the aim to extract generic lessons.
 This contribution is based on Ref.\  \cite{inprep} and accompanied by Ref.\ \cite{denes}.

\section{Formal derivation revisited}
\label{sec:formal}
 
 We will start with revisiting the formal derivation and justification of the complex Langevin approach  \cite{Aarts:2009uq,Aarts:2011ax}  and point out where the arguments have to be amended to include meromorphic drifts. The approach hinges on the equivalence of two expectation values, one defined with respect to the original complex weight $\rho(x)$ and one with respect to the real and positive weight $P(x,y)$ on the analytically extended manifold, i.e.,
\be
\bra O\ket_{\rho(t)} = \int dx\, \rho(x,t)O(x), 
\qquad\quad
\bra O\ket_{P(t)} = \int dxdy\, P(x,y;t)O(x+iy). 
\ee
These distributions satisfy the Fokker-Planck equations (we consider real noise only \cite{Aarts:2009uq})
\be
 \dot\rho(x,t) = \nabla_x\left[\nabla_x -K(x)\right]\rho(x,t),  
 \qquad\quad
 \dot P(x,y;t) =  \left[ \nabla_x\left(\nabla_x -K_x\right) -\nabla_y K_y \right] P(x,y;t),
\ee
with the  Langevin drift terms
\be
 K(z) = \nabla_z\rho(z)/\rho(z), \qquad\quad   K_x = \re\, K(z), \qquad\quad K_y = \im\, K(z).
\ee
Success is obtained when these expectation values are equal: $\bra O\ket_{\rho(t)}  = \bra O\ket_{P(t)}$. 

The equivalence can indeed by demonstrated  \cite{Aarts:2009uq,Aarts:2011ax}, provided that 1) the drift and observables are holomorphic;
2) the distribution $P(x,y)$ has fast decay at $y\to \pm\infty$.
In particular, the proof requires partial integration at $|y|\to\infty$ without boundary terms. 

Let us now consider the case with (at least) one zero in measure, $\rho(x=z_p)=0$. 
In that case the drift  $K(z) = \nabla_z\rho(z)/\rho(z)$ has pole at $z=z_p$ and is no longer holomorphic, but only meromorphic. Hence it is necessary to revisit the derivation.
Note that QCD is an example that falls in this category, since after integrating the quarks, the partition function is
\be
 Z=\int DU\, \det M(U)e^{-S_{\rm YM}}, \qquad\quad
\mbox{with} \;\;\det M(U)=0 \;\;\mbox{for some} \;\; U\in \mbox{SL}(N,\mathbb{C}).
\ee
It turns out that the derivation as above goes through, provided that the region around the pole is excluded, i.e.\ $|z-z_p|>\eps$ \cite{inprep}. However, this yields the possibility of new potential boundary terms at $z\sim z_p$, besides the ones at $|y|\to\infty$. It is therefore necessary to study the behaviour of the product of the distribution and observables, $P(x,y)O(x+iy)$, around $z\sim z_p$ carefully. 
 
Let us make the following remark on the time evolution of holomorphic observables \cite{inprep}, 
\be
 \dot O(z;t) = \tilde L O(z;t), \qquad\qquad \tilde L= \left[\nabla _z +K(z)\right] \nabla_z,
 \ee
with the solution
\be
O(z;t) = e^{\tilde Lt}O(z;0) = \sum_k \frac{t^k}{k!}\tilde L^k O(z;0).
\ee
Since $\tilde L$ has a pole at $z=z_p$, one may expect $O(z;t)$ to have an essential singularity at $z=z_p$. 
However, this potential disaster is counteracted by the vanishing of $P(x,y)$ as $z\to z_p$, as well as by the nontrivial angular dependence of $P(x,y)$ around $z=z_p$ (see below), which soften the singularity.

\section{Poles and the distribution}

\begin{figure}[h]
\centerline{\includegraphics[width=.5\columnwidth]{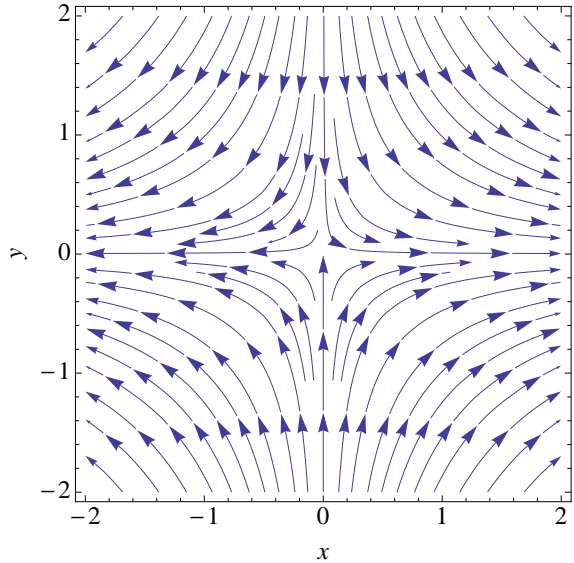}}
\caption{Generic flow pattern around a pole at $z=z_p=0$, with attractive and repulsive directions.}
\label{fig:flowpattern}
\end{figure}

The flow pattern around a pole has a generic structure. Consider a zero at  $z=z_p$ of order $n_p$, 
\be
\rho(x) = (x-z_p)^{n_p} e^{-S(x)}.
\ee
The drift is then given by
\be
K(z) = \frac{\rho'(z)}{\rho(z)} = \frac{n_p}{z-z_p} -S'(z).
\ee
In Fig.\ \ref{fig:flowpattern}, we show the corresponding classical flow pattern, for $z_p=0$ and $S(z)=0$. The attractive and repulsive directions are generic and lead to particular angular dependence. We note that due to this, multiple circlings of the pole are not expected.  
 
  Given the formal justification, it is necessary to better understand the behaviour of the distribution (and observables) around the pole. Logically, there are three possibilities: 1)  the pole is outside the distribution; 2) the pole is on the edge of the distribution; 3) the pole is inside the distribution. We will now encounter these cases in various models.

\section{Pole and distribution in simple model}
 
\begin{figure}[h]
\centerline{\includegraphics[width=.65\columnwidth]{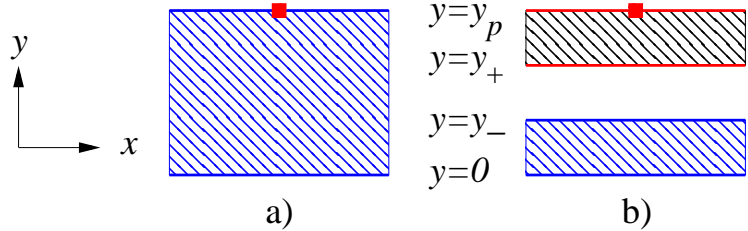}}
 \caption{Strips in the $xy$ plane where $P(x,y)\neq 0$.
 The pole is indicated with the red square. The red striped region in b) is transient only.}
 \label{fig:strip}
\end{figure}

We start with a simple but often studied example, with the distribution
\be
\rho(x) = (x-z_p)^{n_p} e^{-\beta x^2}, \qquad\qquad \beta\in \mathbb{R}, \qquad  z_p=x_p+iy_p\in\mathbb{C}.
\label{eq:strip}
\ee
Following the analysis of Ref.\ \cite{Aarts:2013uza}, it is easy to arrive at some essential and rigorous properties of distribution $P(x,y)$.  For real $\beta$, the distribution is nonzero in a horizontal strip only. Hence the decay at $|y|\to\infty$ poses no problem and this possibility of breakdown is avoided. 
Depending on the parameters ($\beta, y_p, n_p$),  the pole is either  located exactly on the edge of the strip or outside the strip. This is illustrated in Fig.\ \ref{fig:strip}.

These two cases are distinguished by {\cite{inprep}
\bit
\item[a)]  $y_p^2 < 2n_p/\beta$: pole on edge $\Rightarrow P(x,y)\neq 0$ when $0<y<y_p$;
\item[b)]  $y_p^2 > 2n_p/\beta$: pole outside strip $\Rightarrow P(x,y)\neq 0$ when $0<y<y_-<y_p$.
\eit  
For case b), with the pole outside, the standard justification still holds, since the pole is avoided as the upper strip is a transient.
 Indeed, complex Langevin dynamics reproduces the exact results in this case. Case a), with the pole on the edge, is more interesting. Here the results depend on the
properties of the distribution, determined by the parameter values.

\begin{figure}
\centerline{\includegraphics[width=.52\columnwidth]{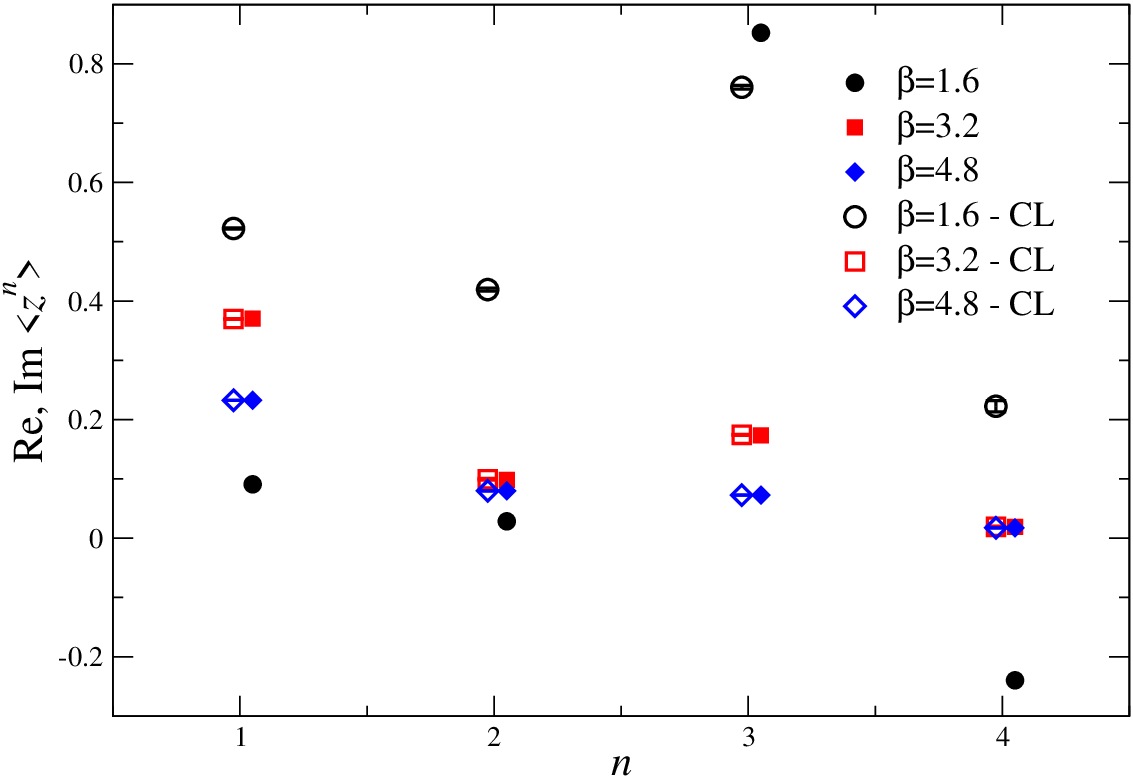}}
   \caption{Observables $\bra z^n\ket$ versus $n$, obtained with complex Langevin (CL, open symbols) and exact results (smaller filled symbols), for three $\beta$ values. 
    }
    \label{fig:zn}
\vspace*{0.2cm}
\centerline{
    \includegraphics[height=5.2cm]{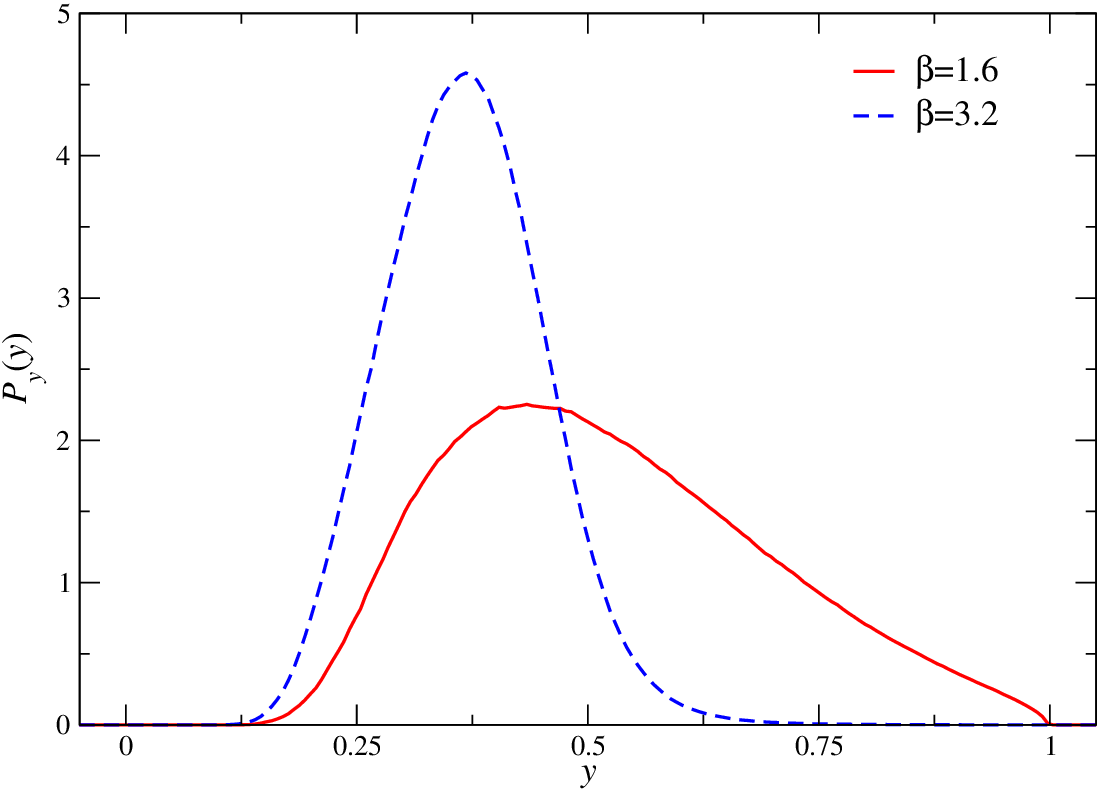}
    \includegraphics[height=5.2cm]{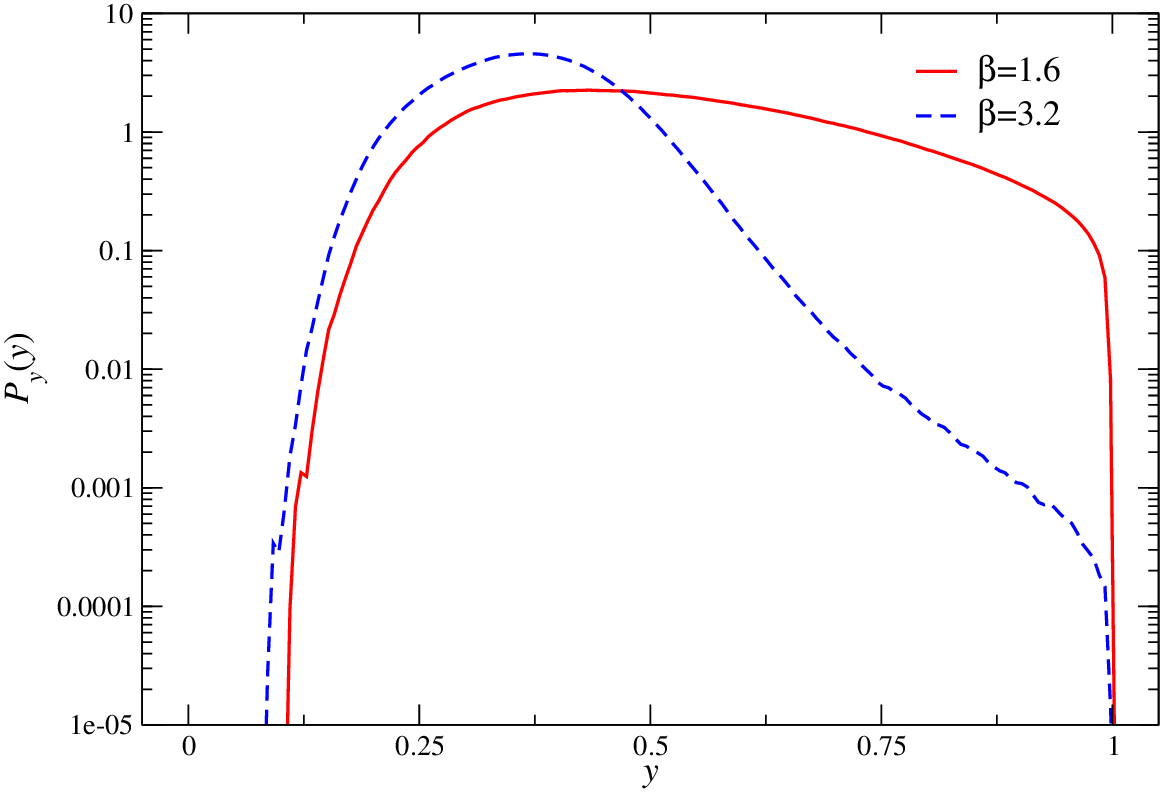}
   }
\caption{Partially integrated distributions $P_y(y)$ on a linear (left) and logarithmic (right) scale.}
\label{fig:pdist}
\end{figure}

 We demonstrate this with an example, using $z_p=i, n_p=2, \beta=1.6, 3.2, 4.8$, and show a comparison of the exact results and the results obtained with CL, for the observable $\bra z^n\ket$, with $n=1,2,3,4$, in Fig.\ \ref{fig:zn}. It can be seen that there is no agreement for $\beta=1.6$, but good agreement for $\beta=3.2$ and $4.8$. Recall that for all three parameter values, $P(x,y)$ is expected to be nonzero for $0<y<y_p=1$, i.e.\ all the way up to the pole. Given the formal justification, this different behaviour should be visible in properties of the distribution $P(x,y)$. 

This is demonstrated in Fig.\ \ref{fig:pdist}, where the partially integrated distribution $P_y(y) = \int dx\, P(x,y)$ is shown on a linear (left) and logarithmic (right) scale, for $\beta=1.6$ (CL incorrect) and 3.2 (CL consistent).
We observe that for $\beta=1.6$, the distribution is nonzero right up to the pole and seems to go to zero linearly. In such a case, boundary terms at $z=z_p$ due to partial integration will contribute and complex Langevin dynamics is not valid, as discussed in Sec.\ \ref{sec:formal}.
On the other hand, for $\beta=3.2$, the decay is much faster, possibly exponentially, and hence partial integration poses no problem for observables $z^n$. Consistent with the formal justification, complex Langevin dynamics then reproduces the correct results.

We conclude that it is possibly to reconcile the properties of the distribution, the formal justification and the success/failure of the complex Langevin process, in the presence of a pole.

\section{Towards more realistic models }
 
The next step is to carry over the essence of this analysis to more realistic models, and devise diagnostics which are also applicable in QCD.
 To do this, we first consider the U(1) one-link model  with the complex distribution \cite{Aarts:2008rr}
\be
\rho(x) = \left[ 1+\kappa\cos(x-i\mu)\right]^{n_p} \exp(\beta\cos x).
\label{eq:weight}
\ee
The conclusions can be summarised (loosely speaking) by stating that CL works when $\kappa<1$ and fails when $\kappa>1$ 
\cite{Aarts:2008rr,Mollgaard:2013qra}. Hence we take here $\kappa=2>1$ here.
We also fix $\beta=0.3, \mu=1$, and vary the order of the zero, $n_p=1,2,4$. 

The distribution has zeroes at $z_p=\pm x_p+i\mu$ and is again nonzero in a strip only, $y_-<y<y_+$, so that the behaviour at $|y|\to \infty$ is under control. We study the observables  $\bra e^{ikz}\ket$  with $k=\pm 1, \pm 2$.
The applicability of CL is found to depend strongly on $n_p$, and we find incorrect results for $n_p=1,2$ but correct results for $n_p=4$ \cite{inprep}.
  In this case, the poles lie within the strip. However, the poles pinch the distribution, i.e.\ approximately disconnected regions appear and the poles act as a bottleneck.  In order to present this in a way that is easily extendable to more complicated theories, where the complexified configuration space is not accessible, we show the determinant factor  $D=1+\kappa\cos(x-i\mu)$ in the complex plane instead. Here $D$ is defined such that the `full determinant' appears as $D^{n_p}$ in Eq.\ (\ref{eq:weight}). 

 \begin{figure}[t]
\centerline{
   \includegraphics[height=5.2cm]{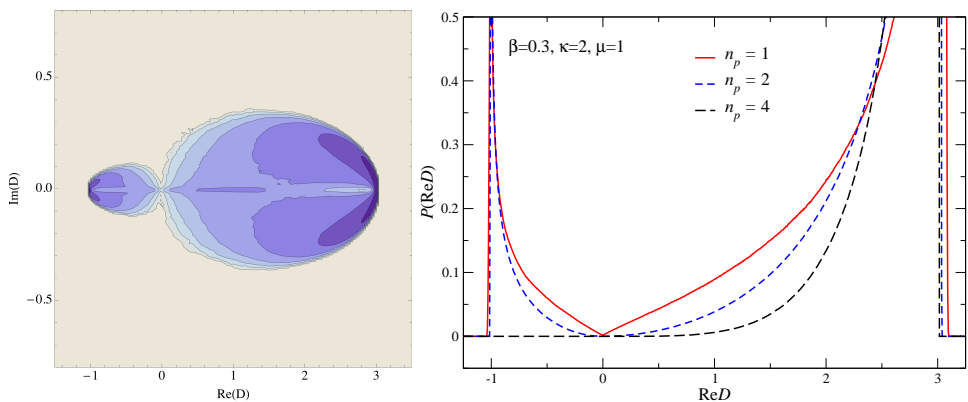}
    \includegraphics[height=5.2cm]{plot-U1-b03-k2-mu1-nf124-PreD.eps}
  }
\caption{Left: histogram of the determinant factor $D$ in the U(1) model for $n_p=2$ (on a logarithmic scale). Right: partially integrated distribution for Re $D$, for three values of $n_p$.   }
\label{fig:u1}
\end{figure}
 
 The result for $n_p=2$ is shown in Fig.\ \ref{fig:u1} (left). We observe that the pole pinches the distribution and, even though the dynamics takes place in the complex plane, there are no multiple circlings around the pole at the origin. The distribution is zero when $D=0$. The way this occurs is shown in Fig.\ \ref{fig:u1} (right) for Re $D$, i.e.\ we have integrated over Im $D$. As in the model discussed above, we note that the manner in which the distribution goes to zero is essential: for $n_p=1, 2$, it is too slow for partial integration to work without boundary terms, while for $n_p=4$, the distribution is in fact zero when Re $D<0$,  partial integration can be applied, and the formal justification is valid.
 
We note that it is now easy to divide the configuration space into two disconnected regions, with Re $D \lessgtr 0$. These regions can be treated separately with constrained partition functions $Z_\pm$ and relative weights $w_\pm = Z_\pm/(Z_++Z_-)$. We find that $Z_-$ typically contributes incorrect results. However, we also find that  $w_-\ll w_+$, which is beneficial for the CL approach.
 
\begin{figure}
\centerline{
    \includegraphics[width=.43\columnwidth]{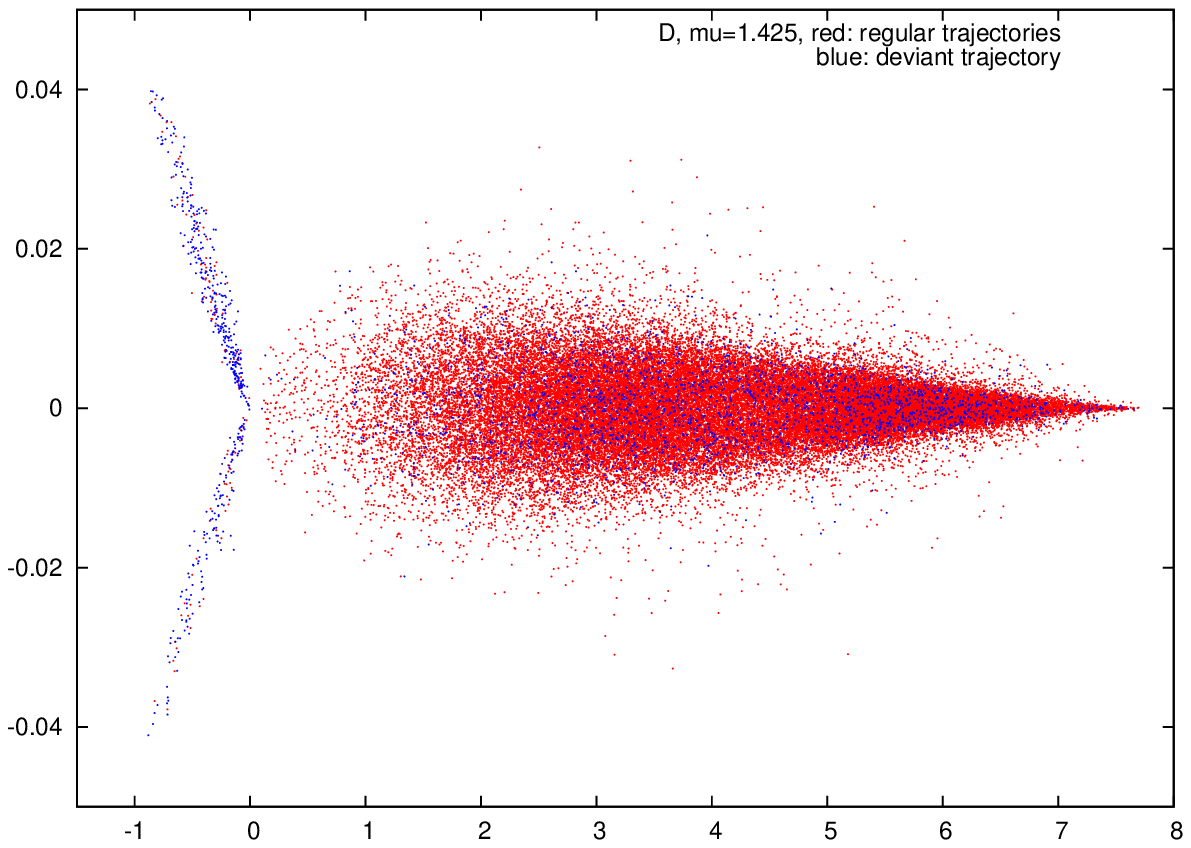}
    \includegraphics[width=.43\columnwidth]{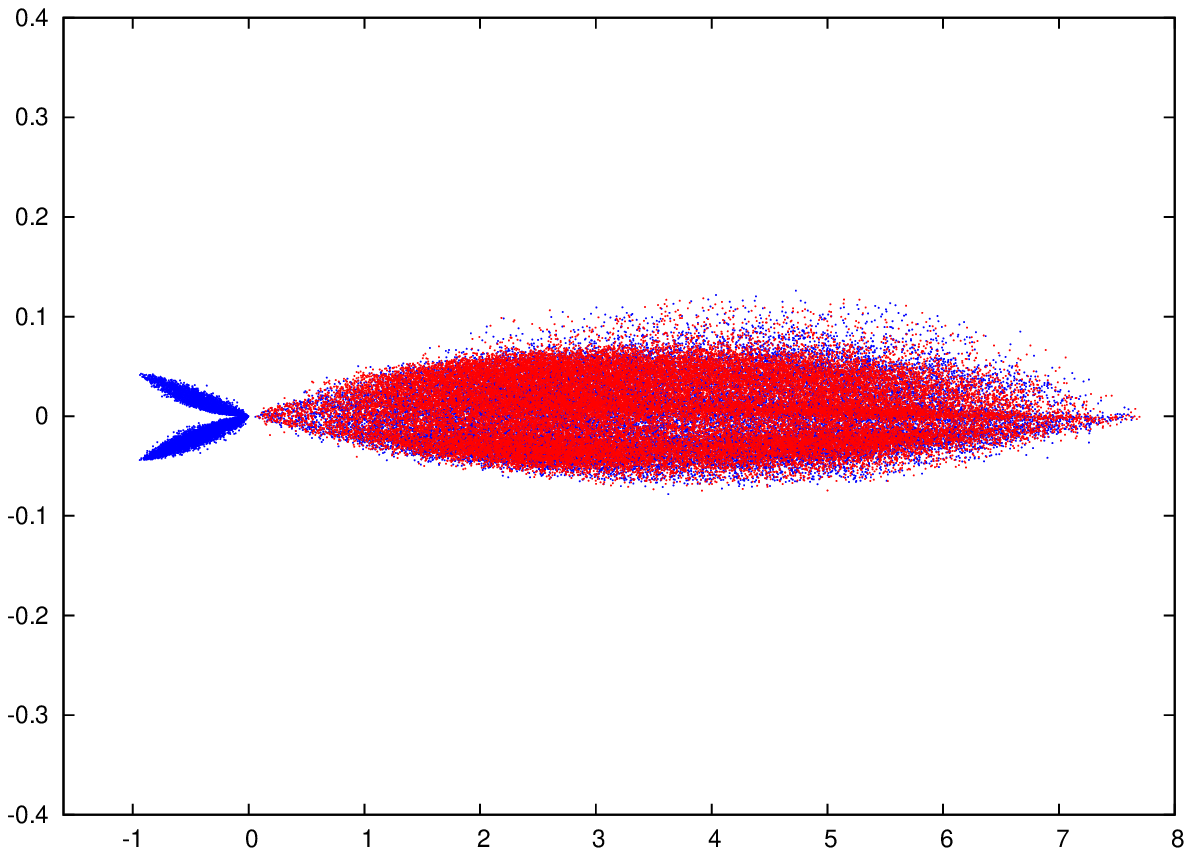}
   }
\caption{Scatter plot of complex determinant in the SU(3) one-link model, for two choices of parameters representing different aspects of HDQCD.
 }
\label{fig:su3}
\end{figure}

Finally we discuss the SU(3) effective one-link model, designed to understand heavy dense QCD \cite{inprep}.
Here it is again straightforward to analyse the determinant and we found the same structure as above, as demonstrated in Fig.\ \ref{fig:su3}.
We find that the zero pinches the distribution, which results in two disjoint areas. It is hence possible to analyse each region separately. When increasing the order of the zero (increasing $n_p$), we find that there is a stronger drift towards and then away from pole, hence  stronger pinching, and typically better agreement with expected results. 
The analysis of determinant is easily extended to heavy dense QCD, while in full QCD it is numerically more intensive. This is further discussed in Refs.\ \cite{inprep,denes}.

 \section{Summary}
 
 When using complex Langevin dynamics to solve QCD at nonzero chemical potential, the Langevin drift has poles where the  fermion determinant vanishes and is no longer holomorphic. It is then necessary to revisit the formal justification of the approach. We found that the usual derivation still holds, but correctness of the results depends crucially on the behaviour of the distribution around the pole.
Similar arguments are also provided in Ref.\ \cite{nagata}.
 Subsequently we analysed  a number of models and found common features in all of these, namely that the poles will pinch the distribution and result in disjoint regions, which can be analysed separately.
When the zero is of order $n_p$, e.g. when the determinant can be written as $[\det D]^{n_p}$, we found that larger $n_p$ typically yields better results. This conclusion seems not specific to simple models, but also correct in e.g. heavy dense QCD \cite{inprep,denes}.

 \vspace*{0.2cm}
 
 \noindent
 {\bf Acknowledgements} --  We thank Felipe Attanasio and Benjamin J\"ager. Support from STFC (grant ST/L000369/1), the Royal Society and the Wolfson Foundation is gratefully acknowledged.


\begin{thebibliography}{99}


\bibitem{Aarts:2013lcm}
  G.~Aarts,
  PoS LATTICE {\bf 2012} (2012) 017
  [arXiv:1302.3028 [hep-lat]].

\bibitem{Sexty:2014dxa}
  D.~Sexty,
  PoS LATTICE {\bf 2014} (2014) 016
  [arXiv:1410.8813 [hep-lat]].

\bibitem{Aarts:2008rr}
  G.~Aarts and I.~O.~Stamatescu,
  JHEP {\bf 0809} (2008) 018
  [arXiv:0807.1597 [hep-lat]].

\bibitem{Seiler:2012wz}
  E.~Seiler, D.~Sexty and I.~O.~Stamatescu,
  Phys.\ Lett.\ B {\bf 723} (2013) 213
  [arXiv:1211.3709 [hep-lat]].



\bibitem{Aarts:2016qrv}
  G.~Aarts, F.~Attanasio, B.~J\"ager and D.~Sexty,
  JHEP {\bf 1609} (2016) 087
  [arXiv:1606.05561 [hep-lat]].

\bibitem{Sexty:2013ica}
  D.~Sexty,
  Phys.\ Lett.\ B {\bf 729} (2014) 108
  [arXiv:1307.7748 [hep-lat]].


\bibitem{Aarts:2014bwa}
  G.~Aarts, E.~Seiler, D.~Sexty and I.~O.~Stamatescu,
  Phys.\ Rev.\ D {\bf 90} (2014) no.11,  114505
  [arXiv:1408.3770 [hep-lat]].

\bibitem{Sinclair:2015kva}
  D.~K.~Sinclair and J.~B.~Kogut,
  PoS LATTICE {\bf 2015} (2016) 153
  [arXiv:1510.06367 [hep-lat]].

\bibitem{Mollgaard:2013qra}
  A.~Mollgaard and K.~Splittorff,
  Phys.\ Rev.\ D {\bf 88} (2013) no.11,  116007
  [arXiv:1309.4335 [hep-lat]].

\bibitem{Splittorff:2014zca}
  K.~Splittorff,
  Phys.\ Rev.\ D {\bf 91} (2015) no.3,  034507
  [arXiv:1412.0502 [hep-lat]].

\bibitem{Nagata:2016alq}
  K.~Nagata, J.~Nishimura and S.~Shimasaki,
  JHEP {\bf 1607} (2016) 073
  [arXiv:1604.07717 [hep-lat]].

\bibitem{Aarts:2009uq}
  G.~Aarts, E.~Seiler and I.~O.~Stamatescu,
  Phys.\ Rev.\ D {\bf 81} (2010) 054508
  [arXiv:0912.3360 [hep-lat]].

\bibitem{Aarts:2011ax}
  G.~Aarts, F.~A.~James, E.~Seiler and I.~O.~Stamatescu,
  Eur.\ Phys.\ J.\ C {\bf 71} (2011) 1756
  [arXiv:1101.3270 [hep-lat]].

\bibitem{inprep}
  G.~Aarts, E.~Seiler, D.~Sexty and I.~O.~Stamatescu, in preparation.
  
\bibitem{denes}
  G.~Aarts, E.~Seiler, D.~Sexty and I.~O.~Stamatescu, 
    PoS(LATTICE2016) 092. 
  
\bibitem{Aarts:2013uza}
  G.~Aarts, P.~Giudice and E.~Seiler,
  Annals Phys.\  {\bf 337} (2013) 238
  [arXiv:1306.3075 [hep-lat]].

\bibitem{nagata}
 K.\ Nagata et al, 
    PoS(LATTICE2016) 067. 





\end{thebibliography}
\end{document}